\documentstyle[aps,prl,preprint,epsfig,graphicx]{revtex}  
\draft  
\tighten 
\begin{document}
{\renewcommand{\thefootnote}{\fnsymbol{footnote}}
\title{
        \begin{flushleft}\includegraphics[width=3cm]{B-logo.epsf}\end{flushleft}
        \vspace{-3cm}
          \vbox{\normalsize%
                \rightline{\hbox{\hfil KEK preprint 2001- 131}}
                \rightline{\hbox{\hfil Belle preprint 2001 - 15}}
           }
        \vspace{3cm}
          Production of Prompt Charmonia in $e^{+}e^{-}$ Annihilation
          at $\sqrt{s} \approx 10.6$~GeV\footnote{submitted to PRL}
        }
\author{The Belle Collaboration}
\maketitle
{\renewcommand{\thefootnote}{\fnsymbol{footnote}}
\begin{center}
  K.~Abe$^{9}$,               
  K.~Abe$^{39}$,              
  T.~Abe$^{40}$,              
  I.~Adachi$^{9}$,            
  Byoung~Sup~Ahn$^{16}$,      
  H.~Aihara$^{41}$,           
  M.~Akatsu$^{21}$,           
  Y.~Asano$^{46}$,            
  T.~Aso$^{45}$,              
  T.~Aushev$^{14}$
% ...
\end{center}

\small
\begin{center}
% ...
$^{9}${High Energy Accelerator Research Organization (KEK), Tsukuba}\\
% ...
$^{14}${Institute for Theoretical and Experimental Physics, Moscow}\\
% ...
$^{16}${Korea University, Seoul}\\
% ...
$^{21}${Nagoya University, Nagoya}\\
% ...
$^{39}${Tohoku Gakuin University, Tagajo}\\
$^{40}${Tohoku University, Sendai}\\
$^{41}${University of Tokyo, Tokyo}\\
% ...
$^{45}${Toyama National College of Maritime Technology, Toyama}\\
$^{46}${University of Tsukuba, Tsukuba}\\
% ...
\end{center}
\normalsize
\begin{abstract}
The production of prompt $J/\psi$, $\psi(2S)$, $\chi_{c1}$ and $\chi_{c2}$ is
studied using a $32.4~{\rm fb}^{-1}$ data sample collected with 
the Belle detector at the $\Upsilon(4S)$ and $60$~MeV below the resonance.
The yield of prompt $J/\psi$ mesons in the $\Upsilon(4S)$ sample is compatible 
with that of continuum production; we set an upper limit  
${\cal B}(\Upsilon(4S) \to J/\psi \, X) < 1.9 \times 10^{-4}$ 
at the $95\%$ confidence level, and find 
$\sigma(e^{+}e^{-} \to J/\psi \, X)=1.47\pm 0.10 \pm 0.13$ pb.
The cross-sections for prompt $\psi(2S)$ and direct $J/\psi$ are measured.
The $J/\psi$ momentum spectrum, production angle distribution and polarization
are studied.
\end{abstract}
%
%
%
%
%
\setcounter{footnote}{0}

The production of prompt charmonia is poorly understood,
and provides an interesting environment to study 
the interplay between perturbative QCD and non-perturbative effects.
A recently developed effective field theory,
called non-relativistic QCD 
(NRQCD)~\cite{bodwin}, provides 
a consistent calculational framework for direct heavy quarkonium production.
Further experimental information is needed to establish the applicability
of NRQCD to charmonium production~\cite{kramer}.
Studies of prompt charmonia in $e^{+}e^{-}$ collisions 
at the $\Upsilon(4S)$ resonance constitute a
test of NRQCD and can provide estimates
of some of its non-perturbative matrix elements.

In this Letter, we report results of a measurement of prompt charmonium 
production in data 
recorded at the $\Upsilon(4S)$ and in the 
continuum $60$ MeV below the resonance, corresponding to integrated
luminosities of $29.4~{\rm fb}^{-1}$ and $3.0~{\rm fb}^{-1}$
respectively~\cite{belleconf}.
The data was collected with the Belle detector 
at the KEKB asymmetric energy ($3.5$~GeV$\,\times\, 8$~GeV)
$e^{+}e^{-}$ storage ring \cite{KEKB}. 

The Belle detector is a large-solid-angle spectrometer equipped
with a 1.5~T superconducting solenoid magnet. 
Charged tracks are reconstructed in a 50 layer Central Drift 
Chamber (CDC)  
and in three concentric layers of double sided silicon strip detectors (SVD).
Photons and electrons are identified using a CsI(Tl) Electromagnetic 
Calorimeter (ECL) located inside the magnet coil.
Muons and $K^{0}_{L}$ mesons are detected using resistive plate
chambers embedded in the iron magnetic flux return (KLM).
Charged particles are identified using specific ionization measurements
in the CDC, pulse heights from the Aerogel \v{C}erenkov 
Counters (ACC) and timing information from the Time of Flight Counters (TOF).
A detailed description of the Belle detector can be found elsewhere \cite{Belle}.

Hadronic events are separated from QED, $\tau\tau$, two-photon and beam-gas
interaction backgrounds by requiring the presence of at least three charged tracks 
($N_{\rm ch}\ge 3$), an event vertex with radial ($r\phi$) and 
$z$ coordinates within 1.5 and 3.5~cm of the origin, respectively,
a total reconstructed center-of-mass (CM) 
energy greater than 0.2$\sqrt{s}$ $(\sqrt{s}$ is the CM collision energy), 
a $z$ component of the net reconstructed CM momentum less than  
0.5$\sqrt{s}/c$, a total ECL energy between 0.1$\sqrt{s}$
and 0.8$\sqrt{s}$ with at least two energy clusters associated, 
and $R_2$, the ratio of second and zeroth Fox-Wolfram moments, 
less than  0.8.

Candidate $J/\psi$ mesons are reconstructed using the leptonic decays
$J/\psi \to \mu^{+}\mu^{-}$ and $e^{+}e^{-}$.
For $J/\psi \to \mu^{+}\mu^{-}$, 
both charged tracks must be identified as muons in the KLM 
using information on hit positions and penetration depth.
For $J/\psi \to e^{+}e^{-}$, oppositely charged track pairs must be
identified as electrons based on a combination of CDC $dE/dx$ information,
ACC response, and the position, shape and energy of the
associated ECL shower.  
To correct for final state radiation and bremsstrahlung,
photons within $50$ mrad of the $e^{\pm}$ are included 
in the $e^{+}e^{-}$ invariant mass calculation.
The two lepton candidate tracks are required to have a common vertex, 
with a distance in the $r\phi$ plane to the average interaction point 
$d_{r\phi} < 500~\mu {\rm m}$.
The signal region is defined by the mass window
$-93 < M_{l^{+}l^{-}}-M_{J/\psi} < 33 ~\rm MeV/c^{2}$,
common for both decay channels.

We reconstruct $\psi(2S) \rightarrow J/\psi \pi^{+}\pi^{-}$ decays by combining
$J/\psi$ candidates with $\pi^+\pi^-$ pairs and requiring a common $\pi^+\pi^-$
vertex.
Candidate $\chi_{c1}$ and $\chi_{c2}$ mesons are reconstructed via their radiative
decays to the $J/\psi$: we combine $J/\psi$ candidates with
photons detected in the ECL, 
that are not associated with an identified $\pi^{0}$.

The largest source of secondary charmonia in the $\Upsilon(4S)$ sample,
due to $B$ meson decays, is eliminated by requiring
the charmonium CM momentum $p^{*}$ to be above the $B$ decay kinematic limit.
A common requirement $p^{*}>2.0$~GeV/$c$ is used for all analyses
($J/\psi$, $\psi(2S)$, $\chi_{c1}$ and $\chi_{c2}$):
this value is robust against the effects of momentum measurement 
errors and motion of the $B$-meson in the CM.
This requirement is not applied to off-resonance data.

The background due to initial state radiation with a hard photon
(``radiative return to $J/\psi\,(\psi(2S))$'')~\cite{eidelman} and higher order QED processes
$e^{+}e^{-}\to J/\psi\, \gamma^{*}$, $J/\psi\, l^{+}l^{-}$~\cite{chang} is also large.
This background contributes mainly to $N_{\rm ch} = 3$ and 
$N_{\rm ch} = 4$ events, and a dedicated study was performed for each of these
samples.
The $J/\psi$ mass peak for $N_{\rm ch} = 3$ has a signal to background ratio
$S/B \approx 0.1$,
which is too poor to be included in the final signal sample.
For $N_{\rm ch} = 4$ the $S/B$ ratio is acceptable, but $85\%$ of the $J/\psi$ mesons
are due to QED processes, principally 
$e^{+}e^{-}\to \psi (2S)\, \gamma \to J/\psi \pi^{+} \pi^{-}\, (\gamma)$ 
with or without a visible photon.
Based on this study we require $N_{\rm ch} > 4$ to suppress the QED backgrounds,
and account for the loss of low multiplicity signal events in the efficiency
calculation.
The remaining QED background in $N_{\rm ch} > 4$ events is suppressed
by rejecting events with a $\psi(2S)\to J/\psi\, \pi^{+}\pi^{-}$ candidate 
accompanied by either a photon with CM energy $E^\ast > 3.5$~GeV  or 
electron tracks from photon conversion.

Mass distributions for prompt $J/\psi$ candidates 
in the $\Upsilon (4S)$ and continuum data samples are shown in Fig.~\ref{mass}. 
Clear signals are observed both in the off-resonance ($226 \pm 23$ events)
and the $\Upsilon (4S)$ data ($1459 \pm 57$ events), for both decay channels.
The average detection efficiency for $J/\psi \to e^+ e^-$ ($\mu^+\mu^-$)
is $38\%$ ($48\%$).

The $\psi (2S)$ signal in $\Upsilon(4S)$ data with $p^{\ast}>2$~GeV/$c$
is shown in Fig.~\ref{massdiff}(a).
Combining $J/\psi \to e^+ e^-$ and $\mu^+ \mu^-$,
a clear signal of $143 \pm 19$ events is seen in the mass difference
$M_{l^{+}l^{-}\pi^{+}\pi^{-}} - M_{l^{+}l^{-}}$ distribution.
In the off-resonance data, no statistically significant $\psi (2S)$ signal is seen ($10 \pm 5$ events);
this is not inconsistent with the observed rate at the $\Upsilon (4S)$ resonance.
Searching for $\chi_{c1}$ and $\chi_{c2}$ in the $\Upsilon (4S)$ data,
we form the mass difference $M_{l^{+}l^{-}\gamma} - M_{l^{+}l^{-}}$;
no significant signals are seen 
(Fig.~\ref{massdiff}(b)).
The average detection efficiency for $\psi (2S)$ and $\chi_{c1}$, $\chi_{c2}$
is $20\%$ and $21\%$ respectively.

We first compare $J/\psi$ production for $p^{*} > 2$~GeV/$c$
in the $\Upsilon (4S)$ and continuum data samples by extracting the number
of signal events from invariant mass distributions 
for $300$~MeV/$c$ wide $p^{*}$ bins.
The shape and normalization of the resulting $p^{*}$ distribution
for the off-resonance continuum (not shown) agree very well with those for the $\Upsilon (4S)$ data,
indicating that all the $J/\psi$ production at the $\Upsilon (4S)$
can be explained by continuum production.
The branching fraction for the decay $\Upsilon (4S) \to J/\psi\, X$
can be estimated from the difference between the normalized yields
($1459 \pm 57$ and $1496 \pm 145$ from the $\Upsilon(4S)$ and off-resonance data respectively)
which is found to be $-37 \pm 156$. 
We therefore set an upper limit 
${\cal B}(\Upsilon(4S) \to J/\psi\, X) < 1.9 \times 10^{-4}$ (at $95\%$ C.L.)
using the Feldman-Cousins method~\cite{feldman}.
This limit is more stringent than the result of Ref.~\cite{babar}.
There are no published predictions for ${\cal B}(\Upsilon(4S) \to J/\psi\, X)$,
but scaling the NRQCD estimate ${\cal B}(\Upsilon(1S) \to J/\psi\, X) \sim 4.5 \times 10^{-4}$~\cite{schuler}
by the ratio of widths $\Gamma_{\Upsilon(1S)}/\Gamma_{\Upsilon(4S)}$ gives $\sim 1.7 \times 10^{-6}$;
this is far below our current experimental sensitivity.

Hereafter we assume that all prompt $J/\psi$'s produced at
$\sqrt{s} \approx 10.6 $~GeV are due to continuum production. 
We combine off-resonance data with the $\Upsilon(4S)$ data for 
$p^* > 2$~GeV/$c$ to extract charmonium production cross-sections
and angular distributions.
Acceptance corrections are determined using a dedicated Monte Carlo (MC)
simulation of the process $e^+ e^- \to J/\psi\,(\psi(2S))\, q \bar{q}$, where 
the composition of $q \bar{q}$ flavors is set to that observed at
$\sqrt{s} \approx 10.6$~GeV.
To minimize model dependence of the efficiency determination, 
the data is corrected using a two-dimensional acceptance weight matrix 
as a function of $p^*$ and $\cos\theta^{*}$, where
$\theta^{*}$ is the $J/\psi$ production angle in the CM system.
The momentum range $p^* > 2$~GeV/$c$ is divided into three bins, and
$\cos\theta^{*}$ into five bins;
the number of signal events in each two-dimensional bin is obtained by 
fitting the invariant mass distribution.
A similar procedure is applied to determine the $\psi(2S)$ acceptance for
$p^{*} > 2$~GeV/$c$, while acceptance corrections for
$\chi_{c1,c2}$ were estimated directly from the MC without binning.

The resulting cross-sections are summarized in Table~\ref{xsections}.
For the $J/\psi$ sample ($p^{*} > 2$~GeV/$c$),
the feed-down from prompt $\psi(2S)$ is measured to be
$0.33 \pm 0.04^{+0.05}_{-0.06}$~pb, and this contribution is subtracted
to give the direct $J/\psi$ cross-section in the table.
The $p^* < 2$~GeV/$c$ off-resonance data is used to extend the 
$J/\psi$ cross-section measurement to the full $p^*$ range.
As we observe no significant signal in the continuum for
$\psi(2S), \chi_{c1}$ and $\chi_{c2}$, we estimate cross-sections for $p^\ast > 2$~GeV/$c$ only.

For the systematic errors of the measured cross-sections the following 
contributions are considered (where appropriate):
uncertainty in the efficiency determination ($\pm 5\%$),
effects of the multiplicity cut ($+ 4\%$),
lepton identification ($\pm 4\%$),
tracking ($\pm 4\%$),
luminosity measurement ($\pm 1.3\%)$,
possible feed-down  from the $\chi_{c1,c2}$ ($- 10\%$),
contamination from QED processes ($- 4\%$),
$\psi(2S)$ feed-down subtraction ($\pm 2\%$)
and errors of the respective branching fractions.
A contribution from the unobserved process $\Upsilon(4S) \to J/\psi\,X$
is not included. 
If the true branching fraction for this decay were just below the determined
upper limit, the measured $J/\psi$ cross-sections would be overestimated
by $\sim 0.18$~pb. 
Other sources of systematic error, such as
contamination from radiative return to the
$\Upsilon (1S,2S,3S)$ and from $\gamma\gamma \to \chi_{c2}$,
were found to be negligible.
As an additional check, $J/\psi$ cross-sections were determined separately for
the $e^+ e^-$ and $\mu^{+} \mu^{-}$ decays, and found to be in good agreement:
their ratio is $1.00 \pm 0.07 \pm 0.04$.

Our results on the $J/\psi$ total and partial cross-section are 
smaller than those of {\sc BaBaR}~\cite{babar}.
The partial cross-sections for direct $J/\psi$ and prompt $\psi(2S)$ production
represent the first measurements in $e^{+}e^{-}$ collisions.

NRQCD predictions for $\sigma(e^+ e^- \to J/\psi_{{\rm direct}}\,X)$
cover the range $0.8 - 1.7$~pb~\cite{yuan}\cite{baek}\cite{schuler}. 
(We exclude the extreme value of Ref.~\cite{chang}.)
The most recent analysis~\cite{schuler} obtains a total cross-section of $0.8 - 1.1$~pb,
attributing $0.3$~pb and $0.5 - 0.8$~pb to color-singlet and color-octet mechanisms respectively.
Our result can be used to further constrain the combination of color-octet matrix
elements needed in the calculation~\cite{yuan}\cite{schuler}.
The ratio
$\sigma (\psi(2S)\,X) / \sigma(J/\psi_{\rm direct}\,X)$ is measured to be
$0.93 \pm 0.11^{+0.13}_{-0.15}$;
this is related in NRQCD to a ratio of linear combinations of non-perturbative 
matrix elements for the respective mesons~\cite{lee}.

Momentum distributions are shown in Fig.~\ref{momentum}.
The distribution of feed-down from $\psi(2S)$ to $J/\psi$ is also shown
in Fig.~\ref{momentum}(a):
it does not significantly modify the overall shape of the $J/\psi$ momentum distribution.
The $J/\psi$ distribution vanishes $\sim 300$~MeV/$c$ 
below the kinematical limit ($4.84$~GeV/$c$)
while the $\psi(2S)$ momentum distribution extends till the end-point ($4.65$~GeV/$c$).
The $J/\psi$ distribution is softer than the NRQCD prediction for
color-singlet $J/\psi\, gg$ \cite{cho}\cite{baek}, and agrees qualitatively with the predicted shape of
the color-singlet $J/\psi\, c\bar{c}$ component \cite{cho}\cite{baek}.
Some NRQCD calculations also predict a dramatic rise in the cross-section at the
end-point due to color-octet $e^+ e^- \to J/\psi\, g$~\cite{braaten}\cite{cho}.
The detection efficiency for this process was studied 
using a dedicated generator embedded in \textsc{Pythia}~\cite{AbeT}.
Assuming a $1$ pb cross-section for this
process, we expect $> 300$ events in the last two bins 
of Fig.~\ref{momentum}(a).
No such signal is observed.

The distributions of the prompt $J/\psi$ CM production angle $\theta^\ast$, and the
helicity angle $\theta_H$ (the angle between 
the positive lepton daughter momentum vector in the
$J/\psi$ rest frame, and the $J/\psi$ momentum vector in the CM system),
have also been studied in different momentum intervals.
We correct the distributions for detection efficiency, but ignore possible
effects due to feed-down. 
The distributions are fitted with
the parameterizations $1 + A\cos^2 \theta^*$ and
$1 + \alpha\cos^2 \theta_{H}$;
the results are summarized in Table~\ref{fits}, with selected fits also shown
in Fig.~\ref{angular}.
No statistically significant $p^{*}$ dependence is seen for either
$A$ nor $\alpha$ parameters.

A large positive $A$ at all momenta for direct $J/\psi$ production is expected
only for the color-singlet 
$J/\psi\, c\bar{c}$ mechanism~\cite{cho}.
However its contribution to 
the cross-section is thought to be small ($\sim 10\%$)~\cite{cho}\cite{baek}.
The leading color octet process $J/\psi\, g$ is expected to yield 
$A \approx +1$ 
at the endpoint~\cite{braaten}\cite{cho}, although as noted above we do not 
observe these events in the $p^{*}$ distribution.
A significant longitudinal polarization of direct $J/\psi$ mesons ($\alpha < -0.4$)
is expected for the color-singlet process $J/\psi\, gg$ alone~\cite{baek}.

In summary, we have observed production of prompt $J/\psi$ and $\psi(2S)$
at energies near the $\Upsilon (4S)$ mass. 
We set an upper limit for $J/\psi$ production from $\Upsilon(4S)$ (valid for $p^{*} > 2$~GeV/$c$),
${\cal B}(\Upsilon (4S) \to J/\psi\, X) < 1.9 \times 10^{-4}$
at the $95\%$ C.L.,
and find the total cross-section for continuum prompt $J/\psi$ production to be
$\sigma (e^+ e^- \to J/\psi\, X) = 1.47 \pm 0.10 ({\rm stat.}) \pm 0.13 ({\rm syst.})$~pb. 
In the momentum range $p^{*} > 2$~GeV/$c$, we measure
$\sigma (e^+ e^- \to J/\psi_{{\rm direct}}\, X) = 0.72 \pm 0.08^{+0.13}_{-0.17}$~pb and
$\sigma (e^+ e^- \to \psi (2S)\, X) = 0.67 \pm 0.09^{+0.09}_{-0.11}$~pb,
and their ratio
$\sigma(\psi(2S)\, X) / \sigma(J/\psi_{{\rm direct}}\, X) = 0.93 \pm 0.17^{+0.13}_{-0.15}$.
The angular distribution of prompt $J/\psi$ mesons 
follows $1 + A\cos^{2}\theta^{*}$ with $A = 0.9 \pm 0.2$,
and the helicity angle follows $1 + \alpha\cos^{2}\theta_H$
with $\alpha = -0.4 \pm 0.1$ indicating 
partial longitudinal polarization.
We do not observe a statistically significant variation of $A$ or $\alpha$ with momentum. 

We wish to thank the KEKB accelerator group for the excellent
operation of the KEKB accelerator. We acknowledge support from the
Ministry of Education, Culture, Sports, Science, and Technology of Japan
and the Japan Society for the Promotion of Science; the Australian
Research
Council and the Australian Department of Industry, Science and
Resources; the Department of Science and Technology of India; the BK21
program of the Ministry of Education of Korea and the CHEP SRC
program of the Korea Science and Engineering Foundation; the Polish
State Committee for Scientific Research under contract No.2P03B 17017;
the Ministry of Science and Technology of Russian Federation; the
National Science Council and the Ministry of Education of Taiwan;
and the U.S. Department of Energy.

\begin{table}[t]
\caption[] {
	The measured charmonium cross-sections.
	The upper limits for $\chi_{c1}$ and $\chi_{c2}$ are at the $90\%$ C.L.}
\begin{tabular}{lcc}
$\sigma$                                    & $0<p^{*}<p^{*}_{max}$         & $2.0<p^{*}<p^{*}_{max}$           \\
   {[pb]}                                   & [GeV/$c$]                     & [GeV/$c$]                         \\
\hline

$\sigma(e^+ e^-\to J/\psi X)$               & $1.47 \pm 0.10 \pm 0.13$      & $1.05 \pm 0.04 \pm 0.09$          \\
$\sigma(e^+ e^-\to J/\psi_{{\rm direct}} X)$&  $-$                          & $0.72 \pm 0.08^{+0.13}_{-0.17}$   \\
$\sigma(e^+ e^-\to \psi(2S) X)$             &  $-$                          & $0.67 \pm 0.09^{+0.09}_{-0.11}$   \\
$\sigma(e^+ e^- \to \chi_{c1} X)$           &  $-$                          & $< 0.35$                          \\
$\sigma(e^+ e^-\to \chi_{c2} X)$            &  $-$                          & $< 0.66$                            
\end{tabular}
\label{xsections}
\end{table}
\begin{table}[t]
\caption[] {
         Results of the fits to angular distributions.}
         
\begin{tabular}{clclc}
$p^{*}_{J/\psi}$~[GeV/$c$] &  \multicolumn{1}{c}{$A$}& $\chi^{2}$/d.o.f.  & \multicolumn{1}{c}{$\alpha$}           & $\chi^{2}$/d.o.f.\\
\hline
$2.0 - 2.6$                &  $0.3^{+0.5}_{-0.4}$    & $1.5/4$              & $-0.4 \pm 0.2$       &  $7.8/4$ \\
$2.6 - 3.4$                &  $1.1^{+0.4}_{-0.3}$    & $5.0/4$              & $-0.4 \pm 0.1$       &  $1.3/4$ \\
$3.4 - 4.9$                &  $1.1^{+0.4}_{-0.3}$    & $4.5/4$              & $-0.2 \pm 0.2$       &  $7.1/4$ \\
\hline
$2.0 - 3.4$                &  $0.7 \pm 0.3$          & $1.2/4$              & $-0.5 \pm 0.1$       &  $4.6/4$ \\ 
$2.0 - 4.9$        &  $0.9 \pm 0.2$                  & $3.0/4$              & $-0.4 \pm 0.1$       &  $13.4/4$ 
\end{tabular}
\label{fits}
\end{table}
\begin{figure}[t]
\centerline{
\epsfxsize 3.2 truein 
\epsfysize 3.0 truein 
\epsfbox{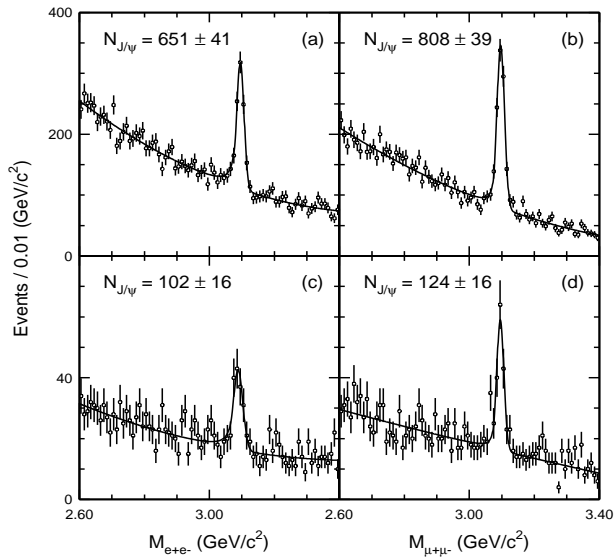}
}
\caption{Mass distributions for the 
        $J/\psi \to e^+ e^-$ (a,c) and $\mu^+ \mu^-$ (b,d) candidates; 
        (a) and (b) are for $\Upsilon (4S)$ data with $p^{*} > 2$~GeV/$c$, 
        (c) and (d) are for off-resonance data. $N_{J/\psi}$ is the number of
	$J/\psi$ mesons determined by a fit to the dilepton mass distribution, 
        where a Crystall Ball line shape function \protect\cite{skwarnicki} is used for the signal 
        and the background is described by a polynomial.
         }
\label{mass}
\end{figure}
\begin{figure}[t]
\centerline{
\epsfxsize 2.8 truein 
\epsfysize 2.8 truein 
\epsfbox{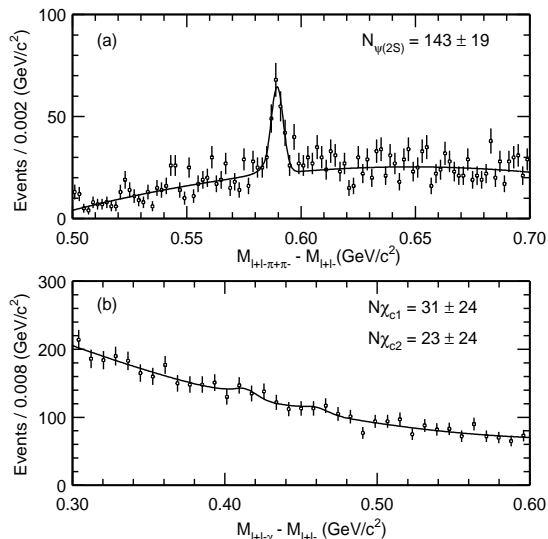}
}
\caption{Mass difference distributions for charmonium candidates in the
	$\Upsilon(4S)$ data: 
         (a) $M_{l^{+}l^{-}\pi^{+}\pi^{-}} - M_{l^{+}l^{-}}$ for $\psi(2S)$ candidates
          with $p^{*}> 2$~GeV/$c$.
         (b) $M_{l^{+}l^{-}\gamma} - M_{l^{+}l^{-}}$ for $\chi_{c1,c2}$ candidates
          with $p^{*}> 2$~GeV/$c$.
          The curves represent fit results with a Crystall Ball line shape for the signal and 
          a polynomial for the background.
           }
\label{massdiff}
\end{figure}
\begin{figure}[t]
\centerline{
\epsfxsize 3.0 truein 
\epsfysize 3.0 truein 
\epsfbox{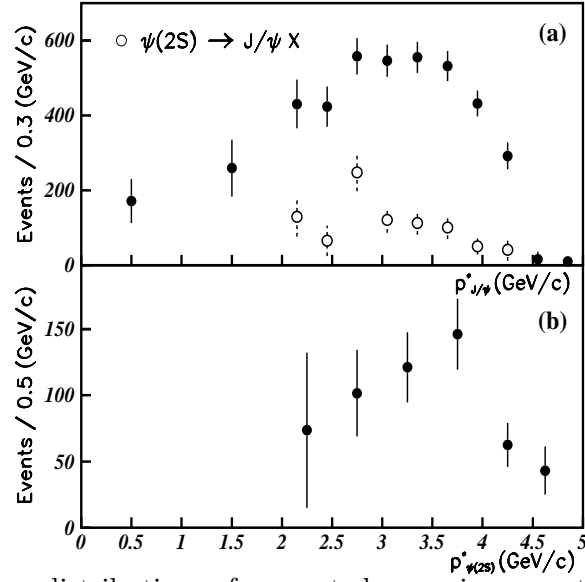}
}   
\caption{CM momentum distributions of prompt charmonia, corrected for efficiency :
         (a) $J/\psi$ (filled points) and $J/\psi$ mesons from $\psi(2S)\to J/\psi\, X$
         (open points);
         (b) $\psi(2S)$.
           }
\label{momentum}
\end{figure}
\begin{figure}[t]
\centerline{
\epsfxsize 2.6 truein 
\epsfysize 2.6 truein 
\epsfbox{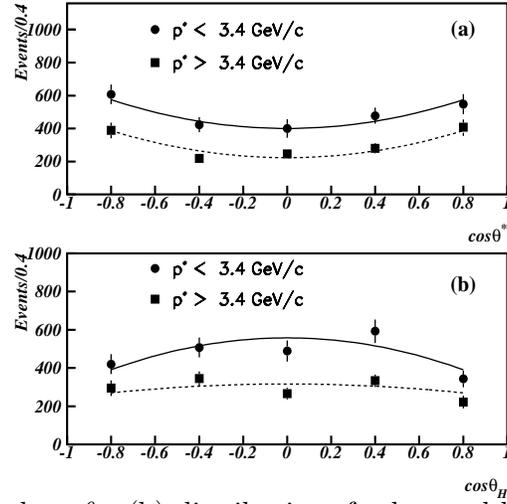}
}   
\caption{The $\cos\theta^{*}$ (a) and  $\cos\theta_{H}$ (b) distributions for
         low and  high $p^{*}$. The curves represent fit results, described in the text and
         summarized in  Table~\ref{fits}.
           }
\label{angular}
\end{figure}
\end{document}